\documentstyle[12pt]{article}
\begin{document}
\renewcommand{\baselinestretch}{1.5}

\centerline{\bf Field Theories of Frustrated Antiferromagnetic}
\centerline{\bf Spin Chains}
\vskip 1 true cm

\centerline{Sumathi Rao{\footnote{{\it e-mail
address} : sumathi@iopb.ernet.in}}$^,$ 
{\footnote{{\it address after Aug. 1$^{st}$} : Mehta Research
Institute, 10 Kasturba Gandhi Marg, \\
\phantom{1111address after Aug.1st : } Allahabad 211002, India}}}
\centerline{\it Institute of Physics, Sachivalaya Marg,}
\centerline{\it Bhubaneswar 751005, India}
\vskip .5 true cm

\centerline{Diptiman Sen\footnote{{\it e-mail address} :
diptiman@cts.iisc.ernet.in}} 
\centerline{\it Centre for Theoretical Studies, Indian Institute
of Science,}
\centerline{\it Bangalore 560012, India}
\vskip 2 true cm

\noindent {\bf Abstract}
\vskip 1 true cm

We study the Heisenberg antiferromagnetic chain with both
dimerization and frustration. The classical ground state has
three phases: a Neel phase, a spiral phase and a colinear phase.
In each phase, we discuss a non-linear sigma model field theory
governing the low energy excitations. We study the theory in the
spiral phase in detail using the renormalization group. The
field theory, based on an $SO(3)$ matrix-valued field, becomes
$SO(3) \times SO(3)$ and Lorentz invariant at long distances
where the elementary excitation is analytically known to be a
massive spin-$1/2$ doublet. The field theory supports $Z_2 ~$
solitons which lead to a double degeneracy in the spectrum for
half-integer spins (when there is no dimerization).

\vskip 1 true cm

\vfill
\eject

Antiferromagnets in low dimensions have been extensively studied
in recent years, partly because of their possible relevance to
high $T_c$ superconductors and partly due to the variety of
theoretical tools which have become available. The latter
include non-linear sigma model (NLSM) field theories [1-6],
Schwinger boson mean field theories \cite{SB}, fermionic mean
field theories \cite{AM}, series expansions \cite{SERIES}, exact
diagonalization of small systems \cite{EXACT}, and the density
matrix renormalization group (DMRG) method \cite{DMRG,CHITRA}.
In one dimension, NLSM theories in particular have received
special attention ever since Haldane \cite{HALDANE} conjectured
that integer spin models would have a gap, contrary to the known
solution for the spin-$1/2$ model, and this prediction was
verified experimentally \cite{HGAPEXPT}.

In this Letter, we study a general Heisenberg spin chain with
both dimerization (an alternation $\delta$ of the nearest
neighbor (nn) couplings) and frustration (a next-nearest
neighbor (nnn) coupling $J_2 ~$). Even classically ($i.e.$, in
the limit where the spin $S \rightarrow \infty$), the system has
a rich ground state `phase diagram', with three distinct phases,
a Neel phase, a spiral phase and a colinear phase (defined
below) \cite{FN1}.  For large but finite $S$, long wavelength
fluctuations about the classical ground state can be described
by non-linear field theories. These field theories are
explicitly known in the Neel phase \cite{HALDANE,AFFLECK} and in
the spiral phase (for $\delta = 0$) \cite{RAO,ALLEN}. While the
Neel phase has been extensively studied, various aspects like
the ground state degeneracy and the low energy spectrum are not
yet well understood in the spiral phase.

We will first discuss the field theory in the Neel phase for
arbitrary $J_2$ and $\delta$. For the spiral phase, we show
using a one-loop renormalization group (RG) analysis that the
field theory flows to an $SO(3) \times SO(3)$ symmetric and
Lorentz invariant theory with an analytically known spectrum
\cite{ANALYTIC}. We also discuss how
the presence of $Z_2$ solitons (supported by the field theory)
affects the ground state degeneracy and the low energy spectrum.
Finally, we show that the field theory in the colinear phase is
qualitatively similar to the one in the Neel phase.

The Hamiltonian for the frustrated and dimerized spin chain is
given by 
\begin{equation}
H ~=~ J_1 ~\sum_i ~(~1~+~(-1)^i \delta ~) {\bf S}_i \cdot {\bf
S}_{i+1} ~ +~ J_2 ~ \sum_i ~{\bf S}_i \cdot {\bf S}_{i+2} ~,
\label{eone}
\end{equation}
where ${\bf S}_i^2 = S(S+1) \hbar^2$, the coupling constants
$J_1, J_2 \ge 0$ and the dimerization parameter $\delta$ lies
between $0$ and $1$. Classically (for $S \rightarrow \infty$),
the ground state is a coplanar configuration of the spins with
energy per spin equal to
\begin{equation}
E_0 ~=~ S^2 ~\left[ {J_1 \over 2} ~ (1+\delta) \cos \theta_1 ~+~
{J_1 \over 2} ~ (1-\delta ) \cos \theta_2 ~+~ J_2 \cos (\theta_1
+ \theta_2) \right]~, 
\end{equation}
where $\theta_1$ is the angle between the spins ${\bf S}_{2i}$
and ${\bf S}_{2i+1}$ and $\theta_2$ is the angle between the
spins ${\bf S}_{2i}$ and ${\bf S}_{2i-1}$.  Minimization of the
classical energy with respect to the $\theta_i ~$ yields the
following phases.

\noindent (i) Neel: This phase has $\theta_1 = \theta_2 = \pi$
and is stable for $1-\delta^2 > 4J_2/J_1$.

\noindent (ii) Spiral: Here, the angles $\theta_1$ and $\theta_2$
are given by
\begin{eqnarray}
\cos \theta_1 &=& - {1\over{1+\delta}}~ {\left[~ {{1 - \delta^2}
\over{4 J_2/J_1}} ~+~ {\delta \over {1 + \delta^2}}~ {4 J_2
\over J_1} ~ \right]} \nonumber\\ {\rm and} \quad
\cos \theta_2 &=& - {1\over{1 -\delta}}~ {\left[~ {{1 -
\delta^2} \over {4 J_2/J_1}} ~-~ {\delta \over {1 - \delta^2}}~
{4 J_2 \over J_1}  ~ \right]},
\end{eqnarray}
where $\pi/2 < \theta_1 <\pi$ and $0<\theta_2 <\pi$. This phase
is stable for $1-\delta^2 < 4 J_2/J_1 < (1-\delta^2) /\delta$.

\noindent (iii) Colinear: This phase (which needs both
dimerization and frustration) is defined to have $\theta_1 =
\pi$ and $\theta_2 = 0$. It is stable for $(1-\delta^2) /\delta
< 4 J_2/J_1$. 

\noindent These phases along with the phase boundaries are
depicted in Fig. 1.

We now study the spin wave spectrum about the ground state
\cite{VILLAIN}.  A detailed analysis will be presented elsewhere
\cite{FUTURE}. We only mention the qualitative results here. In
the Neel phase, we find two zero modes with equal velocities. In
the spiral phase, we have three modes, two with the same
velocity describing out-of-plane fluctuations and one with a
higher velocity describing in-plane fluctuations.  In the
colinear phase, we get two zero modes with equal velocities just
as in the Neel phase.  The distinction between the three phases
is also brought out in the behavior of the spin-spin correlation
function $S(q)$ in the classical limit. $S(q)$ is peaked at
$q=\pi$ in the Neel phase, at $\pi/2 < q < \pi$ in the spiral
phase and at $q=\pi/2$ in the colinear phase.  Even for $S=1/2$
and $1$, DMRG studies have seen this feature of $S(q)$ in the
Neel and spiral phases \cite{CHITRA}. The colinear region has
not yet been probed numerically. 

The spin wave analysis is purely perturbative and is really not
valid since there is no long-range order and no Goldstone modes
in one dimension. To study non-perturbative aspects, we develop
a NLSM to describe the low energy modes. This is well-known in
the Neel phase
\cite{HALDANE,AFFLECK}. The field variable is a unit vector
${\vec \phi}$ and the Lagrangian density is given by
\begin{equation}
{\cal L} ~=~ {(\partial_t {\vec \phi})^2 \over 2 c g^2} ~-~ {c
(\partial_x {\vec \phi})^2\over 2g^2} ~+~ {\theta \over 4 \pi}
{}~{\vec \phi} \cdot \partial_t {\vec \phi} \times \partial_x
{\vec \phi} ~. 
\label{efour}
\end{equation}
Here $c = 2 J_1 aS \sqrt{1-\delta^2 -4 J_2/J_1}~$ is the spin
wave velocity ($a$ is the lattice spacing) and $g^2 = 2/(S
\sqrt{1-\delta^2 - 4 J_2/J_1} )~$ 
is the coupling constant. Note that large $S$ corresponds to
weak coupling. The third term in (4) is a topological term with
$\theta = 2
\pi S (1-\delta)$. This field theory is gapless for $\theta =
\pi$ mod $2\pi$ with the correlation function falling off as a
power at large separations, and is gapped otherwise. For the
gapped theory, the correlations decay exponentially with
correlation length $\zeta$, where $\zeta$ is found from a
one-loop RG calculation to be $\zeta /a =
\exp (2\pi/g^2)$. Hence $\ln (\zeta/a) = \pi S \sqrt{1 -
\delta^2 - 4 J_2/J_1}$. This is plotted in Fig. 2 for $\delta
=0$ and $4 J_2/J_1<1$. 

Recently, the spiral phase has also been studied for $\delta =0$
\cite{RAO,ALLEN}. The classical ground state has $\theta_1 =
\theta_2 = \theta = - J_1/(4 J_2)$. The field variable describing
fluctuations about the classical ground state is an $SO(3)$
matrix ${\underline R}{}(x,t)$ related to the spin variable at
the $i^{\rm th}$ site as $({\bf S}_i)_a = S \sum_b{} {\underline
R}_{ab}{} {\bf n}_b ~$, where $a,b = 1,2,3$ are the components
along the ${\hat {\bf x}}$, ${\hat {\bf y}}$ and ${\hat {\bf
z}}$ axis, and ${\bf n}$ is a unit vector given by
\begin{equation}
{\bf n}_i ~=~ {{{\hat {\bf x}} \cos i\theta + {\hat {\bf y}}
\sin i \theta + a{{\vec \ell}}} \over {\vert ~{\hat {\bf x}}
\cos i\theta + {\hat {\bf y}} \sin i \theta + a{{\vec \ell}}
~\vert}} ~. 
\end{equation}
The unit vector ${\bf n}_i ~$ describes the orientation of the
$i^{\rm th}$ spin in the classical ground state (assumed to lie
in the ${\hat {\bf x}}-{\hat {\bf y}}$ plane) and ${\vec \ell}$
represents the local magnetization. The Hamiltonian in Eq.
(\ref{eone}) can be expanded in terms of ${\underline R}$ and
${{\vec \ell}}$ and Taylor expanded upto second order in
space-time derivatives to obtain a continuum field theory
\cite{ALLEN}. The
Lagrangian density is found to have an $SO(3)_L \times SO(2)_R$
symmetry and can be parametrized as
\begin{equation}
{\cal L} ~=~ {1\over 2c} ~{\rm tr}(\partial_t {\underline R}^T
\partial_t {\underline R} \ P_0) ~-~ {c\over 2} ~{\rm
tr}(\partial_x {\underline R}^T \partial_x {\underline R}\ P_1)~,
\end{equation}
where $c= J_1 S a (1+4 J_2/J_1)\sqrt{1-J_1^2/16 J_2^2}$, and
$P_0$ and $P_1$ are diagonal matrices with entries given by
\begin{eqnarray}
P_0 ~&=&~ \left({1\over 2g_2^2}, ~{1\over 2g_2^2}, ~{{1\over
g_1^2} - {1\over 2 g_2^2}}\right)\nonumber\\
{\rm and} \quad P_1 ~&=&~ \left({1\over 2g_4^2}, ~{1\over
2g_4^2}, {}~{{1\over g_3^2} - {1\over 2g_4^2}}\right) ~.
\end{eqnarray}
The couplings $g_i$ are found to be
\begin{eqnarray}
g_2^2 &=& g_4^2 ~=~ {1\over S} \sqrt{{4J_2+J_1 \over
4J_2-J_1}},\nonumber\\
g_3^2 &=& 2g_2^2, \nonumber\\
{\rm and} \quad g_1^2 &=& g_2^2 ~[~1+(1-J_1 / 2 J_2)^2 ~]~.
\label{eeight}
\end{eqnarray}
Perturbatively, there are three modes, one gapless mode with the
velocity $c g_2/g_4$ and two gapless modes with the velocity $c
g_1/g_3$.  Note that the theory is not Lorentz invariant because
$g_1 g_4 \ne g_2 g_3$. However, the theory is symmetric
under $SO(3)_L \times SO(2)_R$ where the $SO(3)_L$ rotations mix
the rows of the matrix $\underline R$ and the $SO(2)_R$ rotations
mix the first two columns. (To have the full $SO(3)_L \times
SO(3)_R$ symmetry, we need $g_1 = g_2$ and $g_3 = g_4$, $i.e.$,
both $P_0$ and $P_1$ proportional to the identity matrix.) The
$SO(3)_L ~$ is the manifestation in the continuum theory of the
spin symmetry of the original lattice model.  The $SO(2)_R ~$
arises in the field theory because the ground state is planar,
and the two out-of-plane modes are identical and can mix under
an $SO(2)$ rotation. The Lagrangian is also symmetric under the
discrete symmetry parity which transforms ${\underline R}{}(x)
\rightarrow {\underline R}{}(-x) P$ with $P$ being the diagonal
matrix (-1,1,-1).  An important point to note is that there is
no topological term present here (unlike the NLSM in the Neel
phase) and hence, no apparent distinction between integer and
half-integer spins.  There is, however, a distinction due to
solitons, as we will show later.

At distances of the order of the lattice spacing $a$, the values
of the couplings are given in Eq. (8). At larger distance scales
$l$, the effective couplings $g_i(l)$ evolve according to the
$\beta$-functions $\beta(g_i) = d g_i/dy$ where $y = {\rm ln}
(l/a)$. We have computed the one-loop $\beta$-functions using
the background field formalism \cite{SEN}. (Note that since the
theory is not Lorentz-invariant, geometric methods cannot be
used to obtain the $\beta$-functions \cite{AZARIA}.) The
$\beta$-functions are given by
\begin{eqnarray}
\beta(g_1) ~&=&~ {g_1^3 \over {8 \pi}} \left[~{g_1^2 g_3
g_4\over g_2^2} {2\over g_1 g_4 + g_2 g_3} ~+~ 2 g_1 g_3
\ ({1\over g_1^2} - {1\over g_2^2}) ~\right],\nonumber\\
\beta(g_2) ~&=&~ {g_2^3 \over {8 \pi}} \left[~g_1^3 g_3 ({2\over
g_1^2} - {1\over g_2^2})^2 ~+~ 4 g_1 g_3 ~({1\over g_2^2} -
{1\over g_1^2}) ~\right],\nonumber\\
\beta(g_3) ~&=&~ {g_3^3 \over {8 \pi}} \left[~{g_3^2 g_1
g_2\over g_4^2} {2\over g_1 g_4 + g_2 g_3} ~+~ 2 g_1 g_3
\ ({1\over g_3^2} - {1\over g_4^2})~\right],\nonumber\\ {\rm
and} \quad 
\beta(g_4) ~&=&~ {g_4^3 \over {8 \pi}} \left[~g_3^3 g_1 ({2\over
g_3^2} - {1\over g_4^2})^2 ~+~ 4 g_1 g_3 ~({1\over g_4^2} -
{1\over g_3^2})~\right].
\end{eqnarray}
We have numerically investigated the flow of these couplings
using the initial values $g_i(a)$ given in Eq. (8). We find that
the couplings flow such that $g_1/g_2$ and $g_3/g_4$ approach
$1$, $i.e.$, the theory flows towards $SO(3)_L \times SO(3)_R$
and Lorentz invariance. Finally, at some length scale $\zeta$,
the couplings blow up indicating that the system has become
disordered. At one-loop, $\zeta$ depends on $J_2/J_1$ but $S$
can be scaled out. In Fig. 2, we show the numerical results for
$\ln (\zeta /a)$ versus $J_2 / J_1 ~$ for $4 J_2 / J_1 > 1$.
Note that as $4 J_2/J_1 \rightarrow 1$ from either side (the
Neel phase for integer spin or the spiral phase for any spin),
ln($\zeta/a) \rightarrow 0$, $i.e.$, the correlation length goes
through a minimum. Since $4 J_2/J_1 = 1$ separates the Neel and
spiral phases, we may call it a disorder point. (For general
$\delta$, we have a disorder line $4 J_2/J_1 + \delta^2 = 1$ and
the correlation length is minimum on the disorder line
separating the two gapped phases.)

The spiral phase is therefore disordered for any spin $S$ with a
length scale $\zeta$. Since the theory flows to the principal
chiral model with $SO(3)_L \times SO(3)_R$ invariance at long
distances, we can read off its spectrum from the exact solution
given in Ref. \cite{ANALYTIC}. The low energy spectrum consists
of a massive doublet that transforms according to the spin-$1/2$
representation of $SU(2)$. It would be interesting to verify
this by numerical studies of the model. DMRG studies
\cite{DMRG,CHITRA} of spin-$1/2$ and spin-$1$ chains
have not seen these elementary excitations so far. It is likely
that these excitations are created in pairs and a naive
computation of the energy gap would only give the mass of a
pair.  To see them as individual excitations, it would be
necessary to compute the wave function of an excited state and
explicitly compute the local spin density as was done in Ref.
\cite{WHITE} to study a one magnon state in the Neel phase.

Since the field theory is based on an $SO(3)$-valued field
${\underline R}\ (x,t)$ and $\pi_1(SO(3)) = Z_2$, it allows
$Z_2$ solitons.  The classical field configurations come in two
distinct classes with soliton number equal to zero or one. If
${\underline R}_0{} (x,t)$ is a zero soliton configuration, then
a one soliton configuration is obtained as
\begin{equation}
{\underline R}_1{} (x,t) = \left(
         {\begin{array}{ccc}
         \cos\theta (x) & \sin\theta (x) & 0\\
         -\sin\theta (x) & \cos\theta (x) & 0\\
         0 & 0 & 1
         \end{array}}
         \right) {\underline R}_0{} (x,t) ~,
\end{equation}
where $\theta(x)$ goes from $0$ to $2\pi$ as $x$ goes from
$-\infty$ to $+\infty$. (For convenience, we choose $\theta(x) =
2\pi - \theta(-x)$, $i.e.$, the twist is parity symmetric about
the origin.) In terms of spins, this corresponds to
progressively rotating the spins so that the spins at the right
end of the chain are rotated by $2\pi$ with respect to spins at
the left end. Since the derivative $\partial_x \theta$ can be
made vanishingly small, the difference in the energies of the
configurations ${\underline R}_0{} (x,t)$ and ${\underline
R}_1{} (x,t)$ can be made arbitrarily small, and one might
expect to see a double degeneracy in the spectrum.

However, this classical continuum argument needs to be examined
carefully in the context of a quantum lattice model. Firstly, do
${\underline R}_0 \  (x,t)$ and ${\underline R}_1{} (x,t)$
actually correspond to orthogonal quantum states? For the spin
model, if the region of rotation is spread out over an odd
number of sites, $i.e.$, if the rotation operator is $U = {\rm
exp} ({i\pi\over 2m+1} \sum_{n=-m}^m (2n+2m+1) S_n^z)$, then
${\underline R}_0{} (x,t)$ and ${\underline R}_1{} (x,t)$ have
opposite parities because under parity, $S_i^z
\rightarrow - S_i^z$ and $U \rightarrow U {\rm exp}(i2\pi
\sum_{n=-m}^m S_n^z)$. Since the sum contains an odd number of spins,
the term multiplying $U$ is $-1$ for half-integer spin and $1$
for integer spin. Thus for half-integer spin, ${\underline
R}_0{} (x,t)$ and ${\underline R}_1{} (x,t)$ are orthogonal and
the argument for double degeneracy of the spectrum is valid.
This is just a restatement of the Lieb-Schultz-Mattis theorem
\cite{LSM}. For integer spin, ${\underline R}_0 \ (x,t)$ and
${\underline R}_1{} (x,t)$ have the same parity and no
conclusion can be drawn regarding the degeneracy of the
spectrum.

An alternative argument leading to a similar conclusion can be
made following Haldane \cite{FDMH}. We consider a tunneling
process between a zero soliton configuration ${\underline R}_0{}
(x,t)$ and a one soliton configuration ${\underline R}_1{}
(x,t)$. (We choose coplanar configurations for convenience).
Such a tunneling process is not allowed in the continuum theory
(which is why the solitons are topologically stable) because the
configurations have to be smooth at all space-time points. But
in the lattice theory, discontinuities at the level of the
lattice spacing are allowed.  In terms of spins, this tunneling
can be brought about by turning each spin ${\bf S}_i^{(0)}$ in
configuration ${\underline R}_0{} (x,t)$ to the spin ${\bf
S}_i^{(1)}$ in configuration ${\underline R}_1{} (x,t)$ by
either a clockwise or an anticlockwise rotation. Assuming that
the magnitude of the amplitude for the tunneling is the same (as
we will show below), the contribution of the two paths either
add or cancel depending on whether the spin is integral or
half-integral. This is easily seen through a Berry phase
\cite{FRADKIN} calculation.
The difference in the Berry phase of the two paths from ${\bf
S}_i^{(0)}$ to ${\bf S}_i^{(1)}$ is $2 \pi S$. Since the soliton
involves an odd number of spins, the total Berry phase
difference is $0$ mod $2\pi$ if $S$ is an integer and $\pi$ mod
$2\pi$ if $S$ is half-integer.

Now we have to check that the magnitudes of the amplitudes for
tunneling are the same in both the cases. To see this, consider
the pair of spins ${\bf S}_{i}^{(0)}$ and ${\bf S}_{-i}^{(0)}$
which need to be rotated to ${\bf S}_i^{(1)}$ and ${\bf
S}_{-i}^{(1)}$. Since $\theta(x) = 2 \pi -\theta (-x)$, the
magnitude of the amplitude for the clockwise rotation of ${\bf
S}_{i}^{(0)}$ to ${\bf S}_i^{(1)}$ is matched by the magnitude of
the amplitude for the anticlockwise rotation of ${\bf
S}_{-i}^{(0)}$ to ${\bf S}_{-i}^{(1)}$. Hence, for the pair of
spins taken together, the magnitude of the amplitude for
tunneling is the same for the clockwise and anticlockwise
rotations.

Thus, tunneling between soliton sectors is possible for integer
$S$ (thereby breaking the classical degeneracy and leading to a
unique quantum ground state) but not for half-integer $S$ (due
to cancellations between pairs of paths). This agrees with the
earlier Lieb-Schultz-Mattis argument.

Although the NLSM model for the spiral phase was explicitly
derived only for $\delta =0$, we expect the same qualitative
features to persist when $\delta \ne 0$, because the spin wave
analysis shows that the classical ground state continues to be
coplanar and there continue to be three zero modes (two with
identical velocities and the third with a higher velocity
\cite{FUTURE}).
Hence we expect similar RG flows and a similar spectrum.
However, the argument for the double degeneracy of the ground
state for half-integer spins depends on parity being a good
quantum number. When $\delta \ne 0$, parity no longer commutes
with the Hamiltonian and the argument breaks down. This is in
agreement with the DMRG studies \cite{CHITRA} (for periodic
chains) which show a unique ground state, both for integer and
half-integer spins, for $\delta \ne 0$. For open chains, the
ground state is sometimes degenerate due to end degrees of
freedom. To incorporate such effects, one would have to study
NLSM theories on open chains which is beyond the scope of this
work.

Finally, we examine small fluctuations in the colinear phase.
The naive expectation is that the field theory would be an
$O(3)$ NLSM, analogous to the Neel phase, since the classical
ground state is colinear. We can show this explicitly for
$\delta = 1$ which is called the Heisenberg ladder
\cite{LADDER}. The field theory in this limit can be
derived using the classical periodicity under translation by
four lattice sites, similar to the derivation in Ref.
\cite{AFFLECK} for the Neel phase. For two pairs of spins, we
define 
\begin{eqnarray}
\quad {{\vec \phi}} (x) = {{{\bf S}_{4i} - {\bf S}_{4i+1}}\over
2S}, ~ \quad & {{\vec \ell}} (x)& = {{{\bf S}_{4i} + {\bf
S}_{4i+1}}\over 2a}~, \nonumber\\
{\rm and} \quad  {{\vec \phi}} (x)  =  {{\bf S}_{4i+3} - {\bf
S}_{4i+2} \over 2S} ~, &{{\vec \ell}} (x)& = {{\bf S}_{4i+3} +
{\bf S}_{4i+2}\over 2a}~.
\end{eqnarray}
We write the Hamiltonian in terms of the fields ${\vec \phi}$
and ${\vec \ell}$, and Taylor expand to second order in
space-time derivatives to obtain the Lagrangian (\ref{efour})
{\it without} a topological term. We now have
$c=4aS\sqrt{J_2(J_2+J_1)}$ and $g^2 = {1\over S} ~\sqrt{(J_2 +
J_1 )/ J_2}$. The absence of the topological term means that
there is no difference between integer and half-integer spins
and a gap exists in both cases.  In fact, the NLSM predicts a
gap for any finite inter-chain coupling, however small. This is
in agreement with numerical work on coupled spin chains
\cite{LADDER}.

In conclusion, we emphasize that this is the first systematic
field theoretic treatment of the general $J_1-J_2-\delta$ model
on a chain.  It would be interesting to find an experimental
system with sufficient frustration and dimerization to probe the
colinear phase. This phase could also be studied using numerical
techniques like DMRG. The field theoretic treatment of the
spiral phase leads to the interesting possibility that the low
energy excitations of integer spin models may be massive
spin-$1/2$ objects. This again is a possibility which could be
looked for experimentally or verified by numerical simulations.

\vskip 1 true cm

\vfill
\eject

\noindent {\bf Figure Captions}
\vskip 1 true cm

\noindent {1.} Classical phase diagram of the $J_1 -J_2 -\delta$
spin chain.

\noindent {2.} Plot of $\ln (\zeta/a)/S$ versus $J_2/J_1$ for
$\delta =0$. For $4 J_2/J_1<1$, $\ln (\zeta/a)$ is given by the
one-loop RG of the $O(3)$ NLSM for integer spin. For $4
J_2/J_1>1$, $\ln (\zeta/a)$ is given by the one-loop RG of the
$SO(3)_L
\times SO(2)_R$ NLSM.


\begin{thebibliography}{99}

\bibitem{HALDANE} F. D. M. Haldane, Phys. Rev. Lett. {\bf 50}
(1983) 1153; Phys. Lett. A {\bf 93} (1983) 464.

\bibitem{AFFLECK} I. Affleck in {\it Fields, Strings and Critical
Phenomena}, eds. E. Brezin and J. Zinn-Justin (North-Holland,
Amsterdam, 1989); I. Affleck, J.  Phys. Cond. Matt. {\bf 1}
(1989) 3047; Nucl. Phys. B {\bf 257} (1985) 397.

\bibitem{READ} T. Dombre and N. Read, Phys. Rev. B {\bf 39}
(1989) 6797. 

\bibitem{AZARIA} P. Azaria, B. Delamotte and D. Mouhanna, Phys.
Rev. Lett. {\bf 68} (1992) 1762; P. Azaria, B. Delamotte, T.
Jolicouer and D. Mouhanna, Phys. Rev. B {\bf 45} (1992) 12612.

\bibitem{RAO} S. Rao and D. Sen, Nucl. Phys. B {\bf 424} (1994)
547. 

\bibitem{ALLEN} D. Allen and D. Senechal, Phys. Rev. B {\bf 51}
(1995) 6394.

\bibitem{SB} D. P. Arovas and A. Auerbach, Phys. Rev. B {\bf 38}
(1988) 316; D. Yoshioka, J. Phys. Soc. Jpn. {\bf 58} (1989) 32;
S. Sarkar, C. Jayaprakash, H. R. Krishnamurthy and M. Ma, Phys.
Rev. B {\bf 40} (1989) 5028; S. Rao and D. Sen, Phys. Rev. B
{\bf 48} (1993) 12763; R. Chitra, S. Rao, D. Sen and S. S. Rao,
to appear in Phys. Rev. B. 

\bibitem{AM} I. Affleck and J. B. Marston, Phys. Rev. B {\bf 37}
(1988) 3774; J. B. Marston and I. Affleck, Phys. Rev. B {\bf 39}
(1989) 11538; X. G. Wen, F. Wilczek and A. Zee, Phys. Rev. B
{\bf 39} (1989) 11413. 

\bibitem{SERIES} R. R. P. Singh and M. P. Gelfand, Phys. Rev.
Lett. {\bf 61} (1988) 2133.

\bibitem{EXACT} T. Tonegawa and I. Harada, J. Phys. Soc. Jpn.
{\bf 56} (1987) 2153; I. Affleck, D. Gepner, H. J. Schulz and
T. Ziman, J. Phys. A {\bf 22} (1989) 511; K. Okamoto and K.
Nomura, Phys. Lett. A {\bf 169} (1992) 433.

\bibitem{DMRG} S. R. White and D. A. Huse, Phys. Rev. B {\bf 48}
(1993) 3844; S. R. White, Phys. Rev. B {\bf 48} (1993) 10345; Y.
Kato and A. Tanaka, J. Phys. Soc. Jpn. {\bf 63} (1994) 1277.

\bibitem{CHITRA} R. Chitra, S. Pati, H. R. Krishnamurthy, D. Sen
and S. Ramasesha, to appear in Phys. Rev. B; S. Pati, R. Chitra,
D. Sen, H. R. Krishnamurthy and S. Ramasesha, in preparation.

\bibitem{HGAPEXPT} S. H. Glarum, S. Geschwind, K. M. Lee, M. L.
Kaplan and J. Michel, Phys. Rev. Lett. {\bf 67} (1991) 1614; S.
Ma, C. Broholm, D. H. Reich, B. J. Sternlieb and R. W. Erwin,
Phys. Rev. Lett. {\bf 69} (1992) 3571.

\bibitem{FN1} We use the word `phase' for convenience to denote the
position of the peak in the spin-spin correlation function
$S(q)$.  There is actually no phase transition in the spin
chain even at zero temperature.

\bibitem{ANALYTIC} E. Ogievetsky, N. Reshetikhin and P. Wiegmann,
Nucl. Phys. B {\bf 280} (1987) 45.

\bibitem{VILLAIN} J. Villain, J. Phys. (Paris) {\bf 35} (1974) 27.

\bibitem{FUTURE} S. Rao and D. Sen, in preparation.

\bibitem{SEN} For details of the derivation of the
$\beta$-functions, see [5].

\bibitem{WHITE} See S. R. White and D. A. Huse in [11].

\bibitem{LSM} E. H. Lieb, T. Schultz and D. J. Mattis, Ann.
Phys. (NY) {\bf 16} (1961) 407; I. Affleck and E. H. Lieb,
Lett. Math. Phys. {\bf 12} (1986) 57.

\bibitem{FDMH} F. D. M. Haldane, Phys. Rev. Lett. {\bf 61}
(1988) 1029; D. Loss, D. P. DiVincenzo and G. Grinstein, Phys.
Rev. Lett. {\bf 69} (1992) 3232; J. von Delft and C. L. Henley,
{\it ibid.} {\bf 69} (1992) 3236.

\bibitem{FRADKIN} E. Fradkin, {\it Field Theories of Condensed
Matter Systems} (Addison-Wesley, Reading, 1991); E. Manousakis, 
Rev. Mod. Phys. {\bf 63} (1991) 1.

\bibitem{LADDER} S. R. White, R. M. Noack and D. J. Scalapino,
Phys. Rev. Lett. {\bf 73} (1994) 886; T. Barnes, E. Dagotto, J.
Riera and E. S. Swanson, Phys. Rev. B {\bf 47} (1993) 3196; S.
P. Strong and A. J. Millis, Phys. Rev. Lett. {\bf 69} (1992)
2419. 

\end{thebibliography}
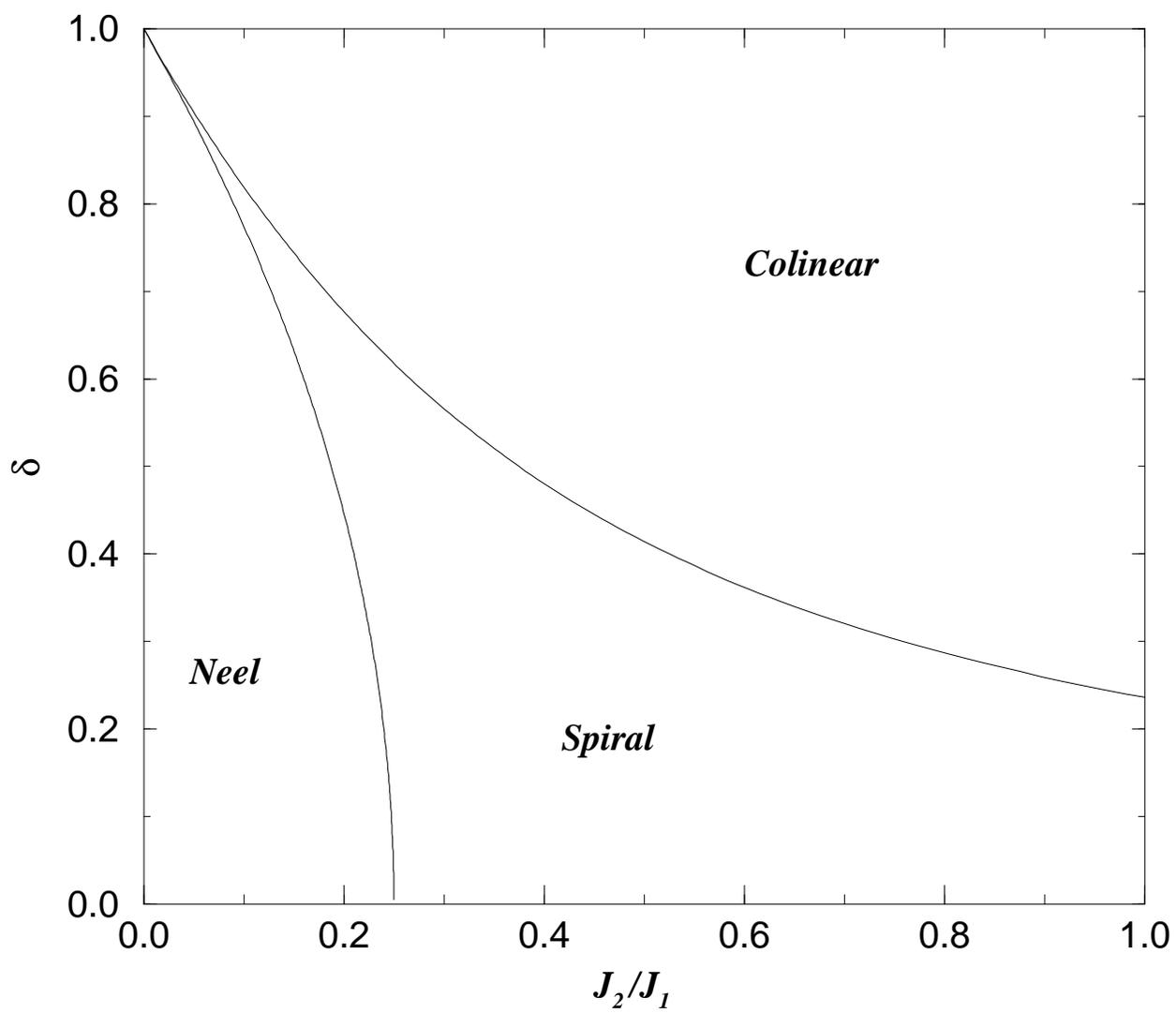
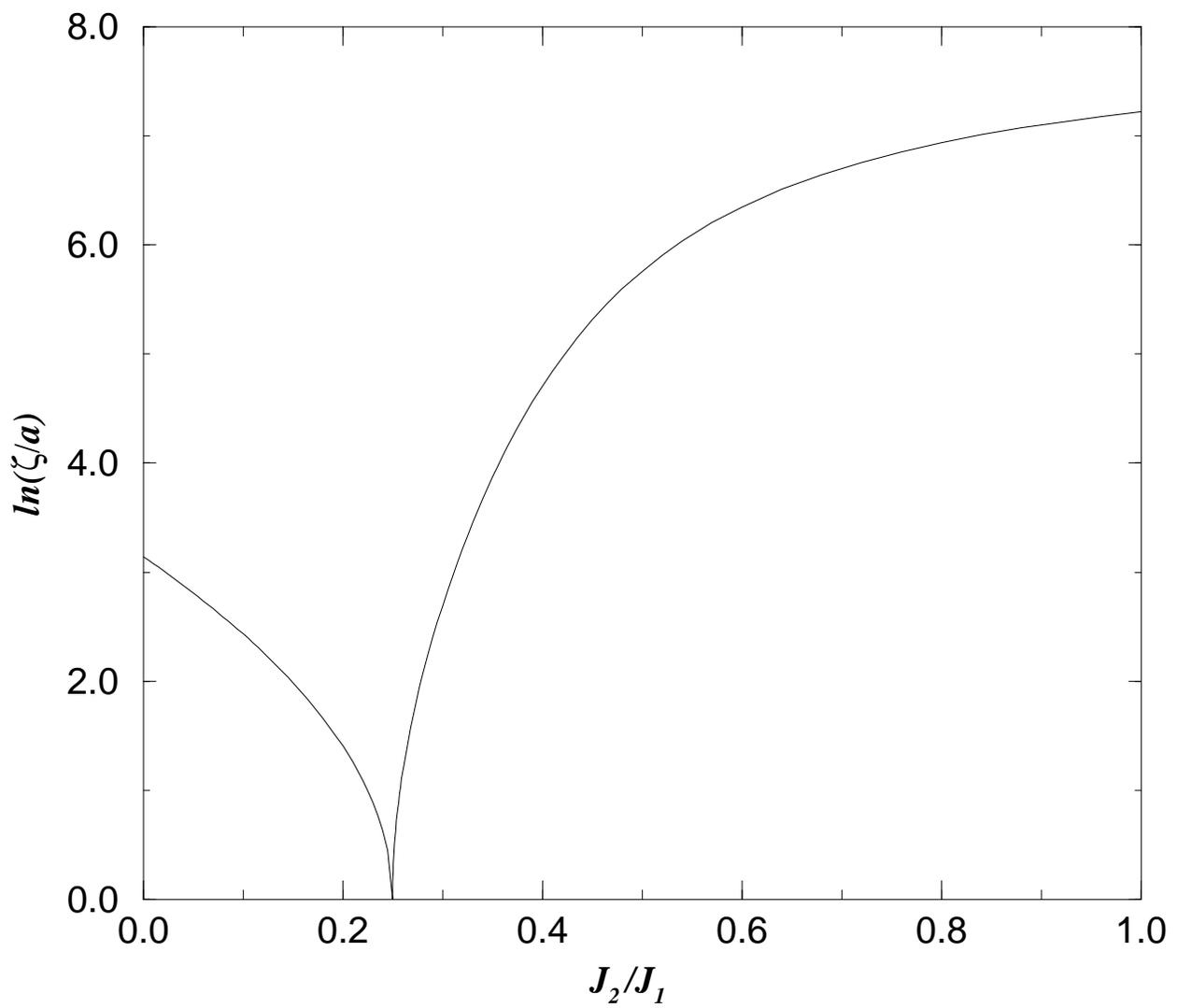
\end{document}